# Computing Nash Equilibria of Action-Graph Games

**Navin A. R. Bhat**
Department of Physics
University of Toronto
Toronto, ON Canada M5S 1A7

**Kevin Leyton-Brown**
Department of Computer Science
University of British Columbia
Vancouver, BC Canada V6T 1Z4

## Abstract

Action-graph games (AGGs) are a fully expressive game representation which can compactly express both strict and context-specific independence between players' utility functions. Actions are represented as nodes in a graph $G$, and the payoff to an agent who chose the action $s$ depends only on the numbers of other agents who chose actions connected to $s$. We present algorithms for computing both symmetric and arbitrary equilibria of AGGs using a continuation method. We analyze the worst-case cost of computing the Jacobian of the payoff function, the exponential-time bottleneck step, and in all cases achieve exponential speedup. When the in-degree of $G$ is bounded by a constant and the game is symmetric, the Jacobian can be computed in polynomial time.

## 1 Introduction

When modeling interactions between self-interested agents in a multiagent system, it is natural to use game theory as a theoretical foundation. (For an introduction to games and equilibrium concepts, see *e.g.*, Fudenberg and Tirole [1991].) The central game-theoretic solution concept is the Nash equilibrium, a fixed-point in mixed strategy space which Nash [1950] proved exists in every finite game. It remains an important open question to determine whether the problem of finding a Nash equilibrium belongs to $\mathcal{P}$ [Papadimitriou, 2001]; the best known algorithms for computing equilibria are exponential. One state of the art general-purpose algorithm is the continuation method of Govindan and Wilson [2003], a gradient-following algorithm which is based on topological insight into the graph of the Nash equilibrium correspondence by Kohlberg and Mertens [1986]. (For a good survey describing earlier algorithms for games with more than two players, see [McKelvey & McLennan, 1996].) The worst-case complexity of Govindan and Wilson's algorithm is open because the worst-case number of gradient-following steps is not known; however, in practice the algorithm's runtime is dominated by the computation of Jacobian of the payoff function, required for computing the gradient, which is both best- and worst-case exponential in the number of agents. For many games this algorithm is an impressive step beyond the previous state of the art; however, it is still only practical when the numbers of players and of actions per player are small (roughly on the order of 5–10).

### 1.1 Compact game representations

One response to the computational difficulty of computing the equilibria of general games has been the investigation of compact game representations that can be leveraged to yield more efficient computation. One influential class of representations exploits strict independencies between players' utility functions; this class includes graphical games [Kearns *et al.*, 2001], multi-agent influence diagrams [Koller & Milch, 2001] and game nets [La Mura, 2000]. Various algorithms were proposed to take advantage of these representations, including exact algorithms for games with special structure [Kearns *et al.*, 2001; Koller & Milch, 2001] and approximate algorithms for arbitrary games [Vickrey & Koller, 2002]. Recently, Blum *et al.* [2003] adapted Govindan and Wilson's continuation algorithm to leverage these representations in an exact algorithm for arbitrary games. Their algorithm computes the Jacobian of the payoff function for unrestricted graphical games and MAIDs. It requires time exponential in the tree width of the underlying graph, an exponential improvement over the Govindan and Wilson algorithm.

A second approach to compactly representing games focuses on *context-specific* independencies in agents' utility functions—that is, games in which agents' abilities to affect each other depend on the actions they choose—and often also on symmetries in agents' utility functions [Rosenthal, 1973; Monderer & Shapley, 1996; Kearns & Mansour, 2002; Roughgarden & Tardos, 2001]. Our past work on *local-effect games* (LEGs) also falls into this class [Leyton-Brown & Tennenholtz, 2003]. LEGs are similar in spirit to action-graph games (AGGs), the game representation we introduce here. They have the same graphical representation, with actions corresponding to nodes, and edges indicating context-specific utility dependencies between actions. However, LEGs involve a series of assumptions: that utility independence between actions is always bidirectional; that all agents share the same action sets; and that utility decomposes into a sum of local effects from individual actions. Thus, LEGs cannot represent arbitrary games. Intuitively, AGGs can be understood as unrestricted LEGs.

## 2 Technical background

### 2.1 Action-graph games

An action-graph game (AGG) is a tuple $\langle N, \mathbf{S}, \nu, u \rangle$. Let $N \equiv \{1, \ldots, n\}$ denote the set of agents in the game. Each agent $i$ has the set of action choices $S_i$, so the set of pure action profiles is

$$\mathbf{S} \equiv \prod_{i \in N} S_i \qquad (1)$$

where $\prod$ is the Cartesian product. Although the actions available to different agents may be distinct, agents may also have action choices in common. Let

$$S \equiv \bigcup_{i \in N} S_i \qquad (2)$$

denote the set of distinct action choices in the game. Let $\Delta$ denote the set of possible distributions of agents over actions, where a distribution is a *number* of agents who chose each action. For a given distribution $D \in \Delta$, denote by $D(s)$ the number of agents who chose action $s$. $\mathcal{D} : \mathbf{S} \mapsto \mathbb{N}^{|S|}$ is a function mapping from a pure strategy profile $\mathbf{s}$ to an agent distribution $D$.

Let $G$ be the action graph: a graph having one node for every action $s \in S$. The neighbor relation is given by $\nu : S \mapsto 2^S$. Let there be a directed edge from $s'$ to $s$ in $G$ iff $s' \in \nu(s)$. Note that $s \in \nu(s)$ is possible. The utility function

$$u : S \times \Delta \mapsto \mathbb{R} \qquad (3)$$

maps from an action choice $s$ and a distribution of agents $D$ to a payoff. Observe that all agents have the same utility function. The utility function has the property that given any action $s$ and any pair of distributions $D$ and $D'$,

$$[\forall s' \in \nu(s),\ D(s') = D'(s')] \Rightarrow u(s, D) = u(s, D'). \qquad (4)$$

In other words, for every $i$ and $j$ agent $i$'s utility is independent of agent $j$'s action choice conditional on agent $j$ choosing an action which is not in the neighborhood of agent $i$'s action choice. This is the sense in which AGGs express context-specific independencies in utility functions. Beyond this condition, there are no restrictions on the function $u$. In some cases it will be notationally simpler for us to write $u(\mathbf{s})$ as a shorthand for $u(s_i, \mathcal{D}(\mathbf{s}))$.

### 2.2 Examples

Any arbitrary game can be encoded as an AGG as follows. Create a unique node $s_i$ for each action available to each agent $i$. Thus $\forall s \in S,\ D(s) \in \{0, 1\}$, and $\forall_1, \sum_{s \in S_i} D(s)$ must equal 1. The distribution simply indicates each agent's action choice, and the representation is no more or less compact than the normal form.

**Example 1** *Figure 1 shows an arbitrary 3-player, 3-action game encoded as an AGG. As always, nodes represent actions and directed edges represent membership in a node's neighborhood. The dotted boxes represent the players' action sets: player 1 has actions 1, 2 and 3; etc. Observe that there is always an edge between pairs of nodes belonging to different action sets, and that there is never an edge between nodes in the same action set.*

In a graphical game [Kearns *et al.*, 2001] nodes denote agents and there is an edge connecting each agent $i$ to each other agent whose actions can affect $i$'s utility. Each agent then has a payoff matrix representing his local game with neighboring agents; this representation is more compact than normal form whenever the graph is not a clique. Graphical games can be represented as AGGs by replacing each node $i$ in the graphical game by a distinct cluster of nodes $S_i$ representing the action set of agent $i$. If the graphical game has an edge from $i$ to $j$, create edges so that $\forall s_i \in S_i, \forall s_j \in S_j,\ s_i \in \nu(s_j)$. The AGG and graphical game representations are equally compact. In Corollary 1 below we show that our general method for computing the payoff Jacobian for AGGs is as efficient as the method specialized to graphical games due to Blum *et al.* [2003].

**Example 2** *Figure 2 shows the AGG representation of a graphical game having three nodes and two edges between them (i.e., player 1 and player 3 do not directly affect each others' payoffs). The AGG may appear more complex than the graphical game; in fact, this is only because players' actions are made explicit.*

The AGG representation becomes even more compact when agents have actions in common, with utility functions depending only on the *number* of agents taking these actions rather than on the *identities* of the agents.

**Example 3** *The action graph in Figure 3 represents a setting in which $n$ ice cream vendors must choose one of four*

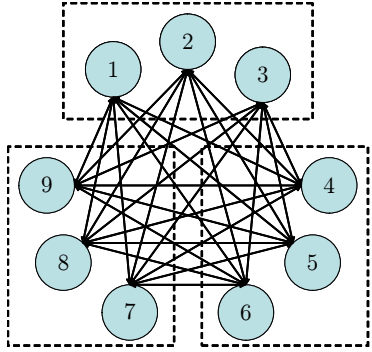
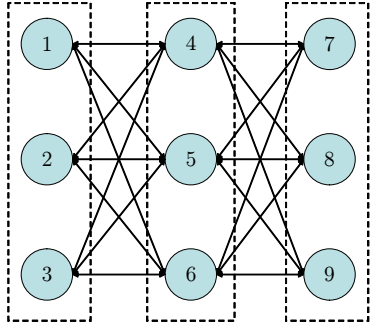
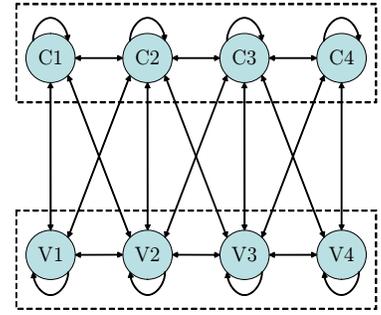

Figure 1: AGG representation of an arbitrary 3-player, 3-action game

Figure 2: AGG representation of a 3-player, 3-action graphical game

Figure 3: AGG representation of the ice cream vendor game

*locations along a beach. Vendors are of two kinds, chocolate and vanilla. Chocolate (vanilla) vendors are negatively affected by the presence of other chocolate (vanilla) vendors in the same or neighboring locations, and are simultaneously positively affected by the presence of nearby vanilla (chocolate) vendors. Note that this game exhibits context-specific independence without any strict independence, and that the graph structure is independent of n.*

Other examples of compact AGGs that cannot be compactly represented as graphical games include: location games, role formation games, traffic routing games, product placement games and party affiliation games.

### 2.3 Notation

Let $\varphi(X)$ denote the set of all probability distributions over a set $X$. Define the set of mixed strategies for $i$ as

$$\Sigma_i \equiv \varphi(S_i), \tag{5}$$

and the set of all mixed strategy profiles as

$$\Sigma \equiv \prod_{i \in N} \Sigma_i. \tag{6}$$

We denote an element of $\Sigma_i$ by $\sigma_i$, an element of $\Sigma$ by $\sigma$, and the probability that player $i$ plays action $s$ by $\sigma_i(s)$.

Next, we give notation for applying some of the concepts defined in Section 2.1 to situations where one or more agents are omitted. By $\Delta_{-\{i,i'\}}$ we denote the set of possible distributions of agents other than $i$ and $i'$, and by $D_{-\{i,i'\}}$ we denote an element of $\Delta_{-\{i,i'\}}$. Analogously, we define $N_{-\{i,i'\}}, S_{-\{i,i'\}}, \Sigma_{-\{i,i'\}}$ and $\sigma_{-\{i,i'\}}$. As a shorthand for the subscript $-\{i,i'\}$, which we will need frequently in the remainder of the paper, we use an overbar, yielding $\overline{\Delta}, \overline{D}, \overline{N}, \overline{S}, \mathbf{S}, \overline{\Sigma}$ and $\overline{\sigma}$. When only one agent is omitted, we write *e.g.* $\Delta_{-i}$. Finally, we overload our notation, denoting by $\mathcal{D}(s_i, s'_i, \overline{D})$ the distribution that results when the actions of $i$ and $i'$ are added to $\overline{D}$.

Define the expected utility to agent $i$ for playing pure strategy $s$, given that all other agents play the mixed strategy profile $\sigma_{-i}$, as

$$V^i_s(\sigma_{-i}) \equiv \sum_{\mathbf{s}_{-i} \in \mathbf{S}_{-i}} u(s, \mathbf{s}_{-i}) \Pr(\mathbf{s}_{-i}|\sigma_{-i}). \tag{7}$$

The set of $i$'s pure strategy best responses to a mixed strategy profile $\sigma_{-i}$ is $\arg\max_s V^i_s(\sigma_{-i})$, and hence the full set of $i$'s pure and mixed strategy best responses to $\sigma_{-i}$ is

$$BR_i(\sigma_{-i}) \equiv \varphi(\arg\max_s V^i_s(\sigma_{-i})). \tag{8}$$

A strategy profile $\sigma$ is a Nash equilibrium iff

$$\forall i \in N, \ \sigma_i \in BR_i(\sigma_{-i}). \tag{9}$$

Finally, we describe the projection of a distribution of agents onto a smaller action space. Intuitively we construct a graph from the point of view of an agent who took a particular action, expressing his indifference between actions that do not affect his chosen action. For every action $s \in S$ define a reduced graph $G^{(s)}$ by including only the nodes $\nu(s)$ and a new node denoted $\emptyset$. The only edges included in $G^{(s)}$ are the directed edges from each of the nodes $\nu(s)$ to the node $s$. The projected distribution $D^{(s)}$ is defined over the nodes of $G^{(s)}$ as

$$D^{(s)}(s') \equiv \begin{cases} D(s') & s' \in \nu(s) \\ \sum_{s'' \notin \nu(s)} D(s'') & s' = \emptyset \end{cases}. \tag{10}$$

In the analogous way, we define $\mathbf{S}^{(s)}, \mathbf{s}^{(s)}, \Sigma^{(s)}$ and $\sigma^{(s)}$.

## 2.4 Continuation Methods

Continuation methods are a technique for solving certain systems of equations when the solution to a perturbed system is known. Given a game of interest with payoff function $u(s_i, \mathbf{s}_{-i})$, one constructs a continuum of games $\Gamma_\lambda$ with payoff functions $u_\lambda$ parameterized by a single parameter $\lambda$, such that $\lambda = 0$ yields the game of interest $\Gamma_0$, and $\lambda = 1$ yields a game $\Gamma_1$ having a known solution. Then beginning with the known solution, the continuation method traces the solution as $\lambda$ is decreased to zero, at which point the desired solution is obtained. A good introduction to continuation methods is given in Blum *et al.* [2003]; we follow their treatment here. A more detailed explanation of the method can be found in Govindan and Wilson [2003].

Nash equilibria are fixed points of a mapping that improves a player's utility by changing his strategy. This mapping yields a system of equations to which continuation methods can be applied. The expected utility of agent $i$ is

$$E[u(s_i, \mathbf{s}_{-i})] = \sum_{s_i \in S_i} \sigma_i(s_i) V^i_{s_i}(\sigma_{-i}), \qquad (11)$$

where we recall that $V^i_{s_i}(\sigma_{-i})$ from Equation (7) is the expected payoff to agent $i$ when he plays pure strategy $s_i$ and other agents play mixed strategy profile $\sigma_{-i}$. Consider the following strategy-improvement mapping for all agents:

$$\sigma' = R(\sigma + V(\sigma)). \qquad (12)$$

Here we introduce the retraction operator $R : \mathbb{R}^m \mapsto \Sigma$. The retraction operator takes a vector of dimension $m \equiv \sum_{i \in N} |S_i|$, and normalizes it by mapping it to the nearest point in $\Sigma$ in Euclidean distance. Mapping from $\sigma$ to $\sigma'$ in Equation (12) corresponds to increasing the probabilities of playing strategies that have better payoffs, while lowering the probabilities of playing strategies that have worse payoffs. Its fixed points $\sigma' = \sigma$, where no further (local) improvement can be achieved for any agent, are the Nash equilibria. Rather than searching in $\Sigma$, Govindan and Wilson found it computationally expedient to search in the unnormalized space $\mathbb{R}^m$ for a $w$ such that $\sigma = R(w)$; then the equilibrium is given by $\sigma = R(w)$ such that $w$ satisfies

$$R(w) = w + R\left(V(R(w))\right). \qquad (13)$$

The perturbed system can now be introduced: replace $V$ with $V + \lambda b$, where $b$ is a bonus to each agent depending on his identity and action choice, but independent of the actions of the other agents. If for each given agent $i$, $b^i_{s_i}$ is sufficiently large for one particular action $s_i$ and zero for all others, then there will be a unique Nash equilibrium where each agent $i$ plays the pure strategy $s_i$. We then have the system of equations $F(w, \lambda) = 0$, where

$$F(w, \lambda) = w - R(w) - (V(R(w)) + \lambda b). \qquad (14)$$

For $\lambda = 1$, $w$ is known; we then wish to decrease $\lambda$ to zero while maintaining $F(w, \lambda) = 0$. This requires that $dF = 0$ along the path. Pairs $(w, \lambda)$ satisfying this condition then map out a graph of the correspondence $w(\lambda)$, which is with probability one over all choices of the bonus a one-manifold without boundary. Thus

$$dF(w, \lambda) = \begin{bmatrix} \nabla_w F & \frac{\partial F}{\partial \lambda} \end{bmatrix} \begin{bmatrix} dw \\ d\lambda \end{bmatrix} = 0 \qquad (15)$$

A nontrivial solution requires that $\begin{bmatrix} \nabla_w F & \frac{\partial F}{\partial \lambda} \end{bmatrix}$ be singular, and its null space defines the direction that is followed by the graph of the correspondence $w(\lambda)$. Using Equation (14) we obtain

$$\nabla_w F = I - (I + \nabla V)\nabla R \qquad (16)$$

where $I$ is the $m \times m$ identity matrix. Computing the Jacobian of $F$ is dominated by the Jacobian of $V$. The derivatives can be taken analytically; elements of the Jacobian for which $i = i'$ vanish, and we obtain for the $i \neq i'$ elements of the Jacobian of $V$,

$$\frac{\partial V^i_{s_i}(\sigma_{-i})}{\partial \sigma_{i'}(s_{i'})} \equiv \nabla V^{i,i'}_{s_i, s_{i'}}(\overline{\sigma}) \qquad (17)$$

$$= \sum_{\overline{\mathbf{s}} \in \overline{\mathbf{S}}} u\left(s_i, \mathcal{D}(s_i, s_{i'}, \overline{\mathbf{s}})\right) Pr(\overline{\mathbf{s}}|\overline{\sigma}) \qquad (18)$$

and

$$Pr(\overline{\mathbf{s}}|\overline{\sigma}) = \prod_{j \in \overline{N}} \overline{\sigma}_j(\overline{\mathbf{s}}_j). \qquad (19)$$

(Recall that whenever we use an overbar in our notation, it is equivalent to the subscript $-\{i, i'\}$. For example, $\overline{\mathbf{s}} \equiv \mathbf{s}_{-\{i,i'\}}$.) Equation (18) shows that the $\nabla V^{i,i'}_{s_i,s_{i'}}(\overline{\sigma})$ element of the Jacobian can be interpreted as the expected utility of agent $i$ when she takes action $s_i$, agent $i'$ takes action $s_{i'}$, and all other agents use mixed strategies according to $\overline{\sigma}$.

The size of the set $\overline{\mathbf{S}}$ is exponential in $n$, but since the sum must visit each member of the set, both best- and worst-case scenario computation of the right-hand side of Equation (18) is exponential in the number of agents. In the sections that follow, we describe algorithms for the efficient computation of this expression.

## 2.5 Other applications of the payoff Jacobian

Efficient computation of the payoff Jacobian is important for more than this continuation method. For example, the iterated polymatrix approximation (IPA) method of Govindan and Wilson [2004] has the same problem at its core. At each step the IPA method constructs a polymatrix game that is a linearization of the current game with respect to the mixed strategy profile, the Lemke-Howson algorithm is used to solve this game, and the result updates the mixed strategy profile used in the next iteration. Though theoretically it offers no convergence guarantee, IPA is typically much faster than the continuation method; often it is used to give the continuation method a quick start. Another application of the payoff Jacobian is in multiagent reinforcement learning algorithms that perform policy search.

## 3 Computing the Jacobian

Whenever $\exists s_i, \exists s'$ such that $s' \notin \nu(s_i)$, there exist distributions of agents that are equivalent from the point of view of agent $i$. Furthermore, whenever $\exists i, \exists j \neq i,\ S_i \cap S_j \neq \emptyset$, there exist pure action profiles that are equivalent from the point of view of agents who choose certain actions. That is, when some action $s$ is available to multiple agents, agents care only about the *number* of agents who choose $s$ and not their identities. We express the first kind of indifference by projecting the action graph; the second is expressed through partitioning the pure action profiles into distributions of agent numbers. Each provides computational benefits. It will be seen below that partitioning amounts to dynamic programming, *i.e.* pushing in of sums that analytically accounts for symmetry in the problem. For arbitrary equilibria, the speedup due to projection is exponential as long as the maximum indegree of the graph $G$ is less than the number of nodes in $G$. This speedup typically overwhelms the gain due to partitioning in non-symmetric equilibria; however, for the important case of symmetric action space (*i.e.* $\forall i, S_i = S$), partitioning guarantees computation of the Jacobian in polynomial time. In Section 3.1 we consider equilibria of arbitrary action-graph games; in Section 3.2 we analyze symmetric equilibria.

### 3.1 Arbitrary equilibria

Given arbitrary action set **S**, the Jacobian can be expressed compactly through projection at the level of the pure action profiles (recall the definition of projection in Equation (10)). Projecting onto action $s_i$, we rewrite the Jacobian given in Equations (18) and (19) as

$$\nabla V_{s_i,s_{i'}}^{i,i'}(\overline{\sigma}) = \sum_{\overline{\mathbf{s}}^{(s_i)} \in \overline{\mathbf{S}}^{(s_i)}} u\left(s_i, \mathcal{D}(s_i, s_{i'}^{(s_i)}, \overline{\mathbf{s}}^{(s_i)})\right) Pr\left(\overline{\mathbf{s}}^{(s_i)} | \overline{\sigma}^{(s_i)}\right)$$
(20)

where

$$Pr\left(\overline{\mathbf{s}}^{(s_i)} | \overline{\sigma}^{(s_i)}\right) = \prod_{j \in \overline{N}} \overline{\sigma}_j^{(s_i)}(\overline{\mathbf{s}}_j^{(s_i)}). \qquad (21)$$

To reflect the indifference between pure action profiles giving rise to the same distribution, we define an equivalence relation that partitions $\overline{\mathbf{S}}$:

$$\overline{\mathbf{s}} \sim \overline{\mathbf{s}}' \text{ iff } \mathcal{D}(\overline{\mathbf{s}}) = \mathcal{D}(\overline{\mathbf{s}}'). \qquad (22)$$

Then denote by $\mathcal{S}(\overline{D})$ the equivalence class containing all $\overline{\mathbf{s}}$ such that $\mathcal{D}(\overline{\mathbf{s}}) = \overline{D}$. The analogous partitioning can be done in the projected space, yielding *e.g.* $\mathcal{S}(\overline{D}^{(s_i)})$.

To exploit this indifference, we rewrite the elements of the Jacobian in Equations (20) and (21) in terms of the projected distribution $\overline{D}^{(s_i)}$ rather than the projected pure action profile $\overline{\mathbf{s}}^{(s_i)}$, obtaining

$$\nabla V_{s_i,s_{i'}}^{i,i'}(\overline{\sigma}) = \sum_{\overline{D}^{(s_i)} \in \overline{\Delta}^{(s_i)}} u\left(s_i, \mathcal{D}\left(s_i, s_{i'}, \overline{D}^{(s_i)}\right)\right) Pr\left(\overline{D}^{(s_i)} | \overline{\sigma}^{(s_i)}\right)$$
(23)

where

$$Pr\left(\overline{D}^{(s_i)} | \overline{\sigma}^{(s_i)}\right) = \sum_{\overline{\mathbf{s}}^{(s_i)} \in \mathcal{S}\left(\overline{D}^{(s_i)}\right)} Pr\left(\overline{\mathbf{s}}^{(s_i)} | \overline{\sigma}^{(s_i)}\right). \qquad (24)$$

Given the value $Pr(\overline{\mathbf{s}}^{(s_i)} | \overline{\sigma}^{(s_i)})$ for some $\overline{\mathbf{s}}^{(s_i)}$, dynamic programming can be used to compute the result for another "adjacent" value $\overline{\mathbf{s}}^{(s_i)'}$ in constant time. We introduce the permutation operation $(j \leftrightarrow j')$, such that $\overline{\mathbf{s}}^{(s_i)'}$ is the pure strategy profile obtained from $\overline{\mathbf{s}}^{(s_i)}$ by switching the actions of agents $j$ and $j'$. Thus $\overline{\mathbf{s}}^{(s_i)'} = \overline{\mathbf{s}}_{(j \leftrightarrow j')}^{(s_i)}$. Then we have that

$$Pr\left(\overline{\mathbf{s}}_{(j \leftrightarrow j')}^{(s_i)} | \overline{\sigma}^{(s_i)}\right) = \frac{\overline{\sigma}_{j'}^{(s_i)}\left(\overline{\mathbf{s}}_j^{(s_i)}\right) \overline{\sigma}_j^{(s_i)}\left(\overline{\mathbf{s}}_{j'}^{(s_i)}\right)}{\overline{\sigma}_j^{(s_i)}\left(\overline{\mathbf{s}}_j^{(s_i)}\right) \overline{\sigma}_{j'}^{(s_i)}\left(\overline{\mathbf{s}}_{j'}^{(s_i)}\right)} Pr\left(\overline{\mathbf{s}}^{(s_i)} | \overline{\sigma}^{(s_i)}\right). \qquad (25)$$

We note that some elements of the Jacobian are identical:

$$\forall s \notin \nu(s_i), s' \notin \nu(s_i), \nabla V_{s_i,s'}^{i,i'}(\overline{\sigma}) = \nabla V_{s_i,s}^{i,i'}(\overline{\sigma}). \qquad (26)$$

Intuitively, Equation (26) expresses context-specific independence: the property that if agent $i'$ takes an action outside the neighborhood of the action taken by $i$, then $i$ is indifferent to which particular action was taken by $i'$. It amounts to projecting the Jacobian. Given that the Jacobian has $O(m^2)$ entries, and given Equation (26), we have that the Jacobian has $O(n^2|S|(\mathcal{I}+1))$ independent entries.

**Theorem 1** *Computation of the Jacobian for arbitrary action-graph games using Equations* (23) *and* (24) *takes time that is* $O\left((\mathcal{I}+1)^{\overline{n}} poly(\overline{n}) poly(|S|)\right)$.

**Proof.** Computing elements of the Jacobian using Equations (23) and (24) involves computing the right hand side of Equation (21) for each $\overline{\mathbf{s}}^{(s_i)} \in \mathcal{S}\left(\overline{D}^{(s_i)}\right)$, for each $\overline{D}^{(s_i)} \in \overline{\Delta}^{(s_i)}$. Since this is just a partitioning of the space $\overline{\mathbf{S}}^{(s_i)}$, it amounts to a sum over all elements of the set $\overline{\mathbf{S}}^{(s_i)}$. This is the same number of elements as in the sum for the Jacobian in Equations (20) and (21). Depending on their specific action sets, each agent has the choice of some subset of the at most $\mathcal{I}+1$ actions in the projected graph $G^{(s_i)}$, and these choices are independent. Thus $\overline{\mathbf{S}}^{(s_i)}$ has $O\left((\mathcal{I}+1)^{\overline{n}}\right)$ elements. Using Equation (26), we have that the Jacobian has $O(n^2|S|(\mathcal{I}+1))$ independent elements. Then computation of the full Jacobian takes time that is $O\left((\mathcal{I}+1)^{\overline{n}} poly(\overline{n}) poly(|S|)\right)$. ∎

**Corollary 1** *For a graphical game encoded as an AGG, if $f$ is the maximum family size and $\alpha$ is the maximum number of actions available to each agent, the Jacobian can be computed in time that is $O\left(poly(\alpha^f) poly(\overline{n}) poly(|S|)\right)$.*

**Proof.** Graphical games can be written as AGGs following the procedure given in Section 2.2. Define the family of agent $i$, $fam(i)$, to be the set of all agents $j$ such that $\exists s_i \in S_i, \exists s_j \in S_j,\ s_j \in \nu(s_i)$. Compute the Jacobian using the method of Equations (20) and (21). The key to the complexity of computing the Jacobian is the size of the set $\overline{\mathbf{S}}^{(s_i)}$. For all elements of $\overline{\mathbf{S}}^{(s_i)}$, all agents $j \notin fam(i)$ take action $\emptyset$. Each agent in $fam(i)$ has at most $\alpha$ available actions in the projected graph $G^{(s_i)}$. Since there are at most $f$ such agents, and since all agents choose independently, therefore $\left|\overline{\mathbf{S}}^{(s_i)}\right| = O(\alpha^f)$. Then the full Jacobian can be computed in time that is $O\left(poly(\alpha^f) poly(\overline{n}) poly(|S|)\right)$. This result is consistent with that of Blum *et al.* [2003], in which the graphical game representation is leveraged to speed up computation of the payoff Jacobian. We note that for strict independence there is a result for AGGs similar to Equation (26) indicating that further elements of the Jacobian are equal; also, the domain of the product in equation (21) can be reduced. These provide further (polynomial) speedup. We omit these results here for reasons of space but note that the full exponential speedup described in [Blum *et al.*, 2003] is already obtained by projecting the action graph. ∎

We note that for an arbitrary action-graph game, it may not be convenient to list the elements of each class $\mathcal{S}\left(\overline{D}^{(s_i)}\right)$; in such a case the Jacobian should be computed using Equations (20) and (21). However, many games have structure that allows convenient iteration over members of these classes, and for such games the method of Equations (23) and (24) provide a constant speedup in computation of the Jacobian. In the method of Equations (20) and (21), the utility function is evaluated $O((\mathcal{I}+1)^{\overline{n}})$ times, whereas using Equations (23) and (24) it is evaluated once for each $\overline{D}^{(s_i)} \in \overline{\Delta}^{(s_i)}$. Consider the operation of extending all agents' action sets via $\forall i, S_i \to S$. Then $\left|\overline{\Delta}^{(s_i)}\right|$ is bounded from above by $\left|\overline{\Delta}^{(s_i)}\right|_{\forall i, S_i \to S}$. This bound is the number of (ordered) combinatorial compositions of $\overline{n}$ into $\left|\overline{S}^{(s_i)}\right|$ nonnegative integers (see *e.g.* [Nijenhuis & Wilf, 1975]),

$$\left|\overline{\Delta}^{(s_i)}\right|_{\forall i, S_i \to S} = \binom{\overline{n} + \left|\overline{S}^{(s_i)}\right| - 1}{\left|\overline{S}^{(s_i)}\right| - 1}. \quad (27)$$

This expression is a polynomial in $\overline{n}$ with degree $\left|\overline{S}^{(s_i)}\right| - 1$, and it follows that

$$\left|\overline{\Delta}^{(s_i)}\right|_{\forall i, S_i \to S} = \Theta\left(\overline{n}^{\left|\overline{S}^{(s_i)}\right| - 1}\right). \quad (28)$$

Thus $\max_{s_i} \left|\overline{\Delta}^{(s_i)}\right| = O(\overline{n}^{\mathcal{I}})$. So whereas computing the Jacobian using Equations (20) and (21) requires evaluating the utility function $O((\mathcal{I}+1)^{\overline{n}})$ times, using Equations (23) and (24) require evaluating the utility function $O(\overline{n}^{\mathcal{I}})$ times. Since $Pr(\overline{\mathbf{s}}^{(s_i)} | \overline{\sigma}^{(s_i)})$ is computed an exponential number of times in both cases, the overall speedup is by a constant factor.

In some games, utility functions depend linearly on the number of agents taking each action. This is true, *e.g.* for local-effect games [Leyton-Brown & Tennenholtz, 2003], where the utility function is defined as

$$u(s_i, D) = \sum_{a \in S} f_{s_i, a}(D(a)). \quad (29)$$

Further dynamic programming is possible in this setting. Given the value of the utility function for one $\overline{D}^{(s_i)}$, it can be evaluated for another $\overline{D}^{(s_i)}_{(a \to a')}$ in constant time using

$$\begin{aligned} u\left(s_i, \overline{D}^{(s_i)}_{(a \to a')}\right) &= u\left(s_i, \overline{D}^{(s_i)}\right) \\ &+ \left(f_{s_i, a}\left(\overline{D}^{(s_i)}_{(a \to a')}(a)\right) - f_{s_i, a}\left(\overline{D}^{(s_i)}(a)\right)\right) \\ &+ \left(f_{s_i, a}\left(\overline{D}^{(s_i)}_{(a \to a')}(a')\right) - f_{s_i, a}\left(\overline{D}^{(s_i)}(a')\right)\right). \end{aligned} \quad (30)$$

Thus the summation in Equation (29) can be evaluated once rather than $|\overline{\Delta}^{(s_i)}|$ times.

### 3.2 Symmetric equilibria

Nash proved [1951] that all finite symmetric games have at least one symmetric equilibrium. We seek such an equilibrium here, specializing to the case in which all agents have the same action choices: $\forall i, \forall j, S_i = S_j = S$. All agents use the same mixed strategy: $\sigma_i = \sigma_j \equiv \sigma_*$. In order to compute a symmetric equilibrium, the continuation method must be seeded with a symmetric equilibrium of the perturbed ($\lambda = 1$) game. This is accomplished by giving all agents the same bonus, so in the perturbed initial equilibrium all agents take the same action. Then since the path-following algorithm is symmetric (*i.e.* the operation of propagation along the path commutes with the permutation of agent identities), the path-following algorithm takes a symmetric perturbed equilibrium to a symmetric equilibrium of the unperturbed game.

Since all agents have the same strategies, each pure action profile is equally likely, so for any $\overline{s} \in \mathcal{S}(\overline{D}^{(s_i)})$

$$Pr\left(\overline{D}^{(s_i)}|\sigma_*^{(s_i)}\right) = \left|\mathcal{S}(\overline{D}^{(s_i)})\right| Pr\left(\overline{s}^{(s_i)}|\sigma_*^{(s_i)}\right), \quad (31)$$

where

$$Pr\left(\overline{s}^{(s_i)}|\sigma_*^{(s_i)}\right) = \prod_{a \in \overline{S}^{(s_i)}} (\sigma_*^{(s_i)}(a))^{\overline{D}^{(s_i)}(a)}. \quad (32)$$

The classes vary exponentially in size. Sizes are given by

$$\left|\mathcal{S}\left(\overline{D}^{(s_i)}\right)\right| = \frac{\overline{n}!}{\prod_{a \in \overline{S}^{(s_i)}} \left(\overline{D}^{(s_i)}(a)\right)!} \quad (33)$$

which is the multinomial coefficient. The largest classes are those in which agents are distributed as uniformly as possible across the nodes of the projected graph (relative to the unprojected graph, this corresponds to having just as many agents choosing each action in $\nu(s_i)$ as choose all the other actions combined).

Furthermore, the Jacobian simplifies, since we need no longer consider individual agent identities in $V$, so instead of considering $\nabla V_{s_i,s_{i'}}^{i,i'}(\overline{\sigma})$, we consider $\nabla V_{*s,s'}(\sigma_*)$, which equals $\nabla V_{s_i,s_{i'}}^{i,i'}(\overline{\sigma})$ for any $i \neq i'$. We replace the strategy improvement mapping of Equation (12) with

$$\sigma_*' = R(\sigma_* + V_*(\sigma_*)). \quad (34)$$

We can thus compute the Jacobian as

$$\nabla V_{*s_i,s_{i'}}(\sigma_*)$$
$$= \sum_{\overline{D}^{(s_i)} \in \overline{\Delta}^{(s_i)}} u\left(s_i, \mathcal{D}\left(s_i, s_{i'}, \overline{D}^{(s_i)}\right)\right) Pr\left(\overline{D}^{(s_i)}|\sigma_*^{(s_i)}\right)$$
(35)

where $Pr(\overline{D}^{(s_i)}|\sigma_*^{(s_i)})$ is given by Eqns (31) and (32). Better still, dynamic programming allows us to avoid reevaluating these equations for every $\overline{D}^{(s_i)} \in \overline{\Delta}^{(s_i)}$. Denote the distribution obtained from $\overline{D}^{(s_i)}$ by decrementing by one the number of agents taking action $a \in \overline{S}^{(s_i)}$ and incrementing by one the number of agents taking action $a' \in \overline{S}^{(s_i)}$ as $\overline{D}^{(s_i)'} \equiv \overline{D}^{(s_i)}_{(a \to a')}$. Then consider the graph $H_{\overline{\Delta}^{(s_i)}}$ whose nodes are the elements of the set $\overline{\Delta}^{(s_i)}$, and whose directed edges indicate the effect of the operation $(a \to a')$. This graph is a regular triangular lattice inscribed within a $(|\overline{S}^{(s_i)}| - 1)$-dimensional simplex. Having computed $Pr(\overline{D}^{(s_i)}|\sigma_*^{(s_i)})$ for one node of $H_{\overline{\Delta}^{(s_i)}}$ corresponding to distribution $\overline{D}^{(s_i)}$, we can compute the result for an adjacent node in $O(|\overline{S}^{(s_i)}|)$ time:

$$Pr\left(\overline{D}^{(s_i)}_{(a \to a')}|\sigma_*^{(s_i)}\right)$$
$$= \frac{\sigma_*^{(s_i)}(a')\overline{D}^{(s_i)}(a)}{\sigma_*^{(s_i)}(a)\left(\overline{D}^{(s_i)}(a') + 1\right)} Pr\left(\overline{D}^{(s_i)}|\sigma_*^{(s_i)}\right). \quad (36)$$

$H_{\overline{\Delta}^{(s_i)}}$ always has a Hamiltonian path [Knuth, unpublished], so having computed $Pr(\overline{D}^{(s_i)}|\sigma_*^{(s_i)})$ for an initial $\overline{D}^{(s_i)}$ using Equation (32), the results for all other projected distributions (nodes) can be computed by using Equation (36) at each subsequent step on the path. Generating the Hamiltonian path corresponds to finding a combinatorial Gray code for compositions; an algorithm with constant amortized running time is given by Klingsberg [1982]. To provide some intuition, it is easy to see that a simple, "lawnmower" Hamiltonian path exists for any lower-dimensional projection of $H_{\overline{\Delta}^{(s_i)}}$, with the only state required to compute the next node in the path being a direction value for each dimension.

**Theorem 2** *Computation of the Jacobian for symmetric action-graph games using Equations* (35), (31), (32) *and* (36) *takes time that is* $O(poly(\overline{n}^{\mathcal{I}})poly(|S|))$.

**Proof.** Recall from Equation (28) that when all agents have the same action choices,

$$\left|\overline{\Delta}^{(s_i)}\right| = \Theta\left(\overline{n}^{|\overline{S}^{(s_i)}|-1}\right). \quad (37)$$

The summation for an element of the Jacobian $V_*$ therefore has $\Theta\left(\overline{n}^{\left|\overline{S}^{(s_i)}\right|-1}\right)$ terms. If the utility function is taken to be straightforward to compute, then the initial summand requires $O(n)$ time. Through dynamic programming the computation of subsequent summands can be done in constant time. Since there are $O(|S|^2)$ independent entries in the Jacobian to be computed, the computation of the symmetric Jacobian takes time $O(poly(\overline{n}^\mathcal{I})poly(|S|))$. ∎

## 4 Conclusions and Future Work

This paper introduced action-graph games, which compactly represent both strict and context-specific independencies between players' utility functions. We showed how the structure of this graph can affect the computation of the Jacobian of the payoff function, the bottleneck step of the Govindan and Wilson continuation algorithm. We presented algorithms for computing both general and symmetric equilibria of AGGs. We showed that in the general case, computation of the Jacobian grows exponentially with the action graph's maximal in-degree rather than with its total number of nodes, yielding exponential savings. We also showed that the Jacobian can be computed in polynomial time for symmetric AGGs when the action graph's maximal in-degree is constant, and described two dynamic programming techniques to further speed up this case.

The full version of this paper will include two sections that could not be included here. First, a game is $k$-symmetric if agents have one of $k$ types and all agents of a given type affect other agents in the same way (see *e.g.* Example 3). Nash's proof that a symmetric equilibrium exists in every finite symmetric game implies that a $k$-symmetric equilibrium exists in every finite $k$-symmetric game. We will extend our results to the $k$-symmetric case, showing that the Jacobian can still be computed in polynomial time for constant $k$. Second, we will provide an implementation and experimental evaluation of our algorithms, derived from the publicly available implementation of Blum *et al.* [2003].